# Image Protection against Forgery and Pixel Tampering based on a Triple Hybrid Security Approach

Ahmed M. Negm, Mohamed Torky*, Mohammed Mahmoud Abo Ghazala, and Hosam Eldin Fawzan Sayed

*Scientific Research Group in Egypt (SRGE)
ahmadnagm@gmail.com
*mtorky86@gmail.com
http://www.egyptscience.net/

**Abstract.** Due to the widespread of advanced digital imaging devices, forgery of digital images became more serious attack patterns. In this attack scenario, the attacker tries to manipulate the digital image to conceal some meaningful information of the genuine image for malicious purposes. This leads to increase security interest about protecting images against integrity tampers. This paper proposes a novel technique for protecting colored images against forgery and pixel tamper. The proposed approach is designed as a hybrid model from three security techniques, Message Digest hashing algorithm (MD5), Advanced Encryption Standard-128 bits (AES), and Stenography. The proposed approach has been evaluated using set of image quality metrics for testing the impact of embedding the protection code on image quality. The evaluation results proved that protecting image based on Least Significant Bit (LSB) is the best technique that keep image quality compared with other two bit-substitution methods. Moreover, the results proved the superiority of the proposed approach compared with other technique in the literature.

## 1 Introduction

Data forgery, the crime of falsely and fraudulently altering or tampering data content represent a classical problem since many decades . With the rapid technological advances in representing and manipulating data in different forms such as texts, images, videos, or sounds make the digital forgery is an important problem [1]. The ability to create digital photographs opened up the doors for producing forged images for malicious purposes. Due to the vast using of digital images across Internet , social media , TV channels, electronic newspapers and magazine , the attackers can execute variety of image forgery attack patterns for malicious purposes such as misusing the reputation of individuals, companies, and countries using fake profiles attacks across social media [2] [3]. Digital check forgery attack [4] is another attack scenario by which, the attacker uses some digital image processing techniques for forging the financial activities across online banking. A Copy-Move forgery attack [5] is a another attack scenario of image tampering where a piece of the image is copied and pasted on another part to hide unwanted portions of the image. The attackers goal is covering the truth or to enhance the visual effect of the image for deceiving purposes. Hence, authenticating digital images and insuring its integrity against pixel tampering and forgery represents a major security problem [6][7].

The literature on detecting digital image forgery has highlighted several techniques and approaches. Copy-move and Splicing image forgery detection [8][9] is a common method for detecting images forgery based on dividing digital image into blocks and use block-matching algorithms for finding the similarities between blocks. The similar blocks will yield similar features. Hence, forgery detection decision is made only if similar features are detected within the same distance of features associated to connected blocks [10]. Discrete Cosine Transform (DCT) [11] is a good example of copy-move detection mechanism which has the ability to detect tampered regions of pixels accurately. Another example of copy-move forgery detection is the Scale-Invariant Feature Transform (SIFT) [12] which is a feature detection technique for detecting and describing local features in the digital images. Discreet Wavelet Transformation and un decimated dyadic wavelet transform (DyWT) [13] are another techniques that can be used to transform image pixels into wavelets, which are then used for wavelet-based compression and coding for detecting copy-move pixel tampering or alteration.

Watermarking [14][15][16][17] can be used as a popular means for efficient image tamper detection. The watermarking-based forgery detection is executed by marking small blocks of an image with watermarks that depend on a secret ID of that particular digital camera and later verify the presence of those watermarks for detecting image forgery.

In this paper, hybrid approach is proposed for detecting digital image forgery and pixels tampering. The proposed model is designed based on three security techniques, Message Digest hashing algorithm (MD5), Advanced Encryption Standard-128 bits (AES), and Steganography. The rest of this paper can be organized as follows: section 2 presents and discusses the proposed model for detecting image forgery. Section 3 presents the experimental results. Section 4 discusses and compare obtained results. Finally, section 5 is devoted to the conclusion of this study.

## 2      Triple Hybrid Security Approach

A number of techniques have been developed in the literature for addressing the problem of protecting images against pixel tampering and forgery [18]. One of the most well-known techniques is watermarking-based algorithms [19][20]. However, watermarking- based techniques are vulnerable against some attack models such as copy attacks, Ambiguity attacks, collision attacks, and scrambling attacks, etc [21]. Other robust forgery detection methods is based on Speed up Robust

Features (SRF) [22][23] which partly inspired by the Scale-Invariant Feature Transform (SIFT) [24] . However these methods still suffer from some advanced geometric attacks such as one pixel attack [25].

In this study, we try to introduce a novel security approach for securing images against advanced forgery attacks such as one pixel attack. The proposed approach integrated three security techniques, MD5 hashing, AES-128 bit, and steganography to produce a triple hybrid security model for protecting images against forgery and pixel tampering. For achieving this objective, the proposed approach is designed based on three security techniques, MD5 hashing [26], AES-128 bit [27] and steganography [28].

1. MD5 hashing based on SHA-160 is used as a secure hash function for protecting image integrity against altering and modification by malicious adversary [29]. It used here for converting the camera ID code into a fixed length-160 bit hash code called Secret Originality Identifier (SOI).
2. AES-128 is a symmetric block cipher based on 128- bits. AES is utilized in this work for encrypting the input colored image (only green and blue Matrices after XOR operation) using the first 16 bits of SOI as a key and produces a cipher matrix. 4*4 pixels.
3. Steganography is a hiding technique used for concealing a file, message, image, or video within another file, message, image, or video. In this work the steganography is used as follows: , the cipher matrix is substituted with the red matrix for producing a modified matrix. Then, the modified matrix is demosaicated with the blue and green matrices for producing the protected RGB image. Three substitution techniques are performed , LSB (Least Significant Bit), MSB (Most Significant Bit), and Fourth Bit (#4 bit).

The methodology of the proposed approach is explained in Figure 1. The technique can be conceived through major five stages:

1. The MD5-SHA160 hash function is used for converting the camera ID code (16 bits) into a fixed length hash code called Secret Originality Identifier (SOI) (16 bits). The main advantage of MD5 is that it is a secure cryptographic hash technique for protecting data integrity and able to detect unintentional data tampering efficiently.
2. The captured image is filtered using the common Color Filter Array (CFA) for specifying input image into RGB architecture in the form Red, Green, and Blue matrices.
3. The Green Matrix (GM) is XORed with the Blue matrix (BM) into a new matrix which then encrypted using AES algorithm and SOI as the encryption key (16-bits) for producing a Cipher Matrix (CM) for the input image.
4. The Cipher Matrix (CM) then substituted with the Red matrix (RM) pixel by pixel in three forms of substitutions for each byte for producing a Modified Matrix (MM). Least Significant Bit (LSB) substitution, Most Significant Bit substitution, and 4-bit substitution are the three substitution techniques applied in this work.

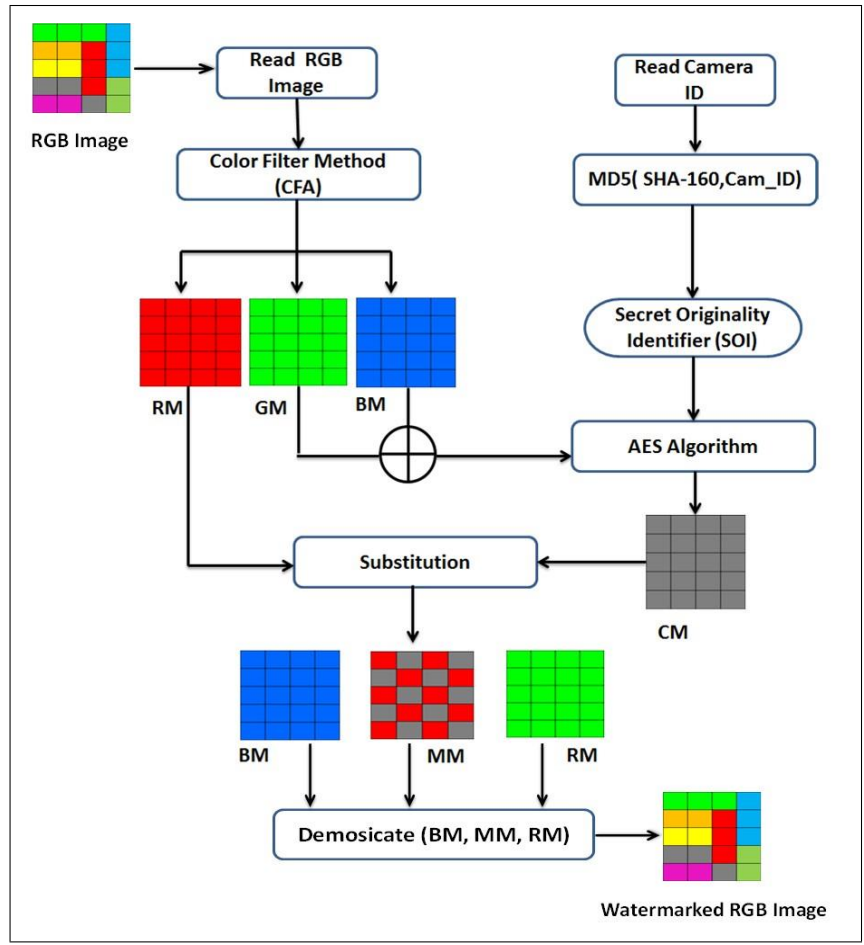

Fig. 1: Image protection based on a triple hybrid security Approach

5. The Modified Matrix (MM) then demosaicated with the Blue matrix (BM) and Green Matrix (GM) for producing the protected Watermarked RGB image again.

## 3    Experimental results

For evaluating the proposed image protection technique,we simulated it using MATLAB 8.5 on Five colored images from the dataset MCC-F220 [30]. Table 1 depicts the image samples and their description. Three experiments-based on three bit substitution methods (least Significant Bit (LSB), 4-Significant Bit (4-SB), and Most Significant Bit (MSB)) are performed on the five images for getting the best RGB watermarked . We used five metrics for testing the efficiency of the proposed technique: Mean Absolute Error (MAE) , Mean Square Error (MSE), Peak Signal to Noise Ratio (PSNR), Structural Similarity Index (SSIM) , and Universal Image Quality Index (UIQI)[31] as depicted in equations 1, 2, 3, 4, 5 respectively.

| Image | Image Name | Size in Pixels | Types | Size in KB |
|---|---|---|---|---|
| 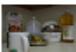 | Mug | 800*532 | JPEG | 31KB |
| 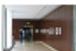 | Lane | 800*532 | JPEG | 33.7KB |
| 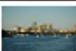 | Sea | 800*532 | JPEG | 57KB |
| 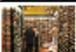 | Library | 737*492 | JPEG | 93KB |
| 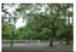 | Garden | 800*532 | JPEG | 123KB |

Fig. 2: Selected Images for simulation experiment

$$MAE = (\frac{1}{N}) \times \sum_{i=1}^{N} |x_i - v_i|) \qquad (1)$$

where $N$ is the number of column vectors, and $v_i$ are variables of paired observations, where Examples of $x_i$ versus $v_i$ include comparisons of predicted versus observed.

$$MSE = (\frac{1}{M*N}) \times \sum_{i=1}^{m} \sum_{j=1}^{n} [x(i,j) - v(i,j)]^2) \qquad (2)$$

Where $M$ is the number of Row vectors, $N$ is the number of column vectors, $x(i, j)$ and $v(i, j)$ are variables of paired observations at pixel (i,j).

$$PSNR = 10 \times \log_{10}(\frac{R^2}{MSE}) \qquad (3)$$

where $R$ is the maximum possible pixel value of the image which is 255, and $MSE$ is the Mean Square Error.

$$SSIM(x, v) = (2 \times x \times v + c1) \times \frac{2 \times \delta_{xv} + c2}{(x^2 + v^2 + c1) \times (\delta_x^2 + \delta_v^2 + c2)} \qquad (4)$$

where, $x$ is the average of $x$. $v$ is the average of $v$, $\delta_x^2$ is the variance of $x$, $\delta_v^2$ is the variance of $v$, $\delta_{xv}$ is the covariance of $x$ and $v$, c1 and c2 are two variables to stabilize the division with weak denominator.

$$UIQ = \left(\frac{\delta_{xy}}{\delta_x \times \delta_y}\right) \times \left(\frac{2\bar{x}\bar{y}}{\bar{x}^2 + \bar{y}^2}\right) \times \left(\frac{2\times\delta_x\times\delta_y}{\delta_x^2 + \delta_y^2}\right) \tag{5}$$

Tables 1,2,3 provide the simulation results based on three bit substitution methods: least Significant Bit (LSB), 4-Significant Bit (4-SB), and Most Significant Bit (MSB). The results proves that least Significant Bit (LSB) is the best bit-substitution method that can be used with the proposed approach for protecting colored images against forgery and pixel tampering.

Table 1: Image Quality Measurements Based on Least Significant Bit (LSB)

| Images/Size/ JPEG | MAE | MSE | PSNR | SSIM | UIQI |
|---|---|---|---|---|---|
| Mug/800*532, 31KB | 0.0842 | 0.165982587 | 55.93017831 | 0.998701529 | 0.999811542 |
| Lane/800*532, 33.7KB | 0.0832 | 0.166969838 | 55.90442334 | 0.998094754 | 0.999707432 |
| Sea/800*532, 57KB | 0.0856 | 0.166371269 | 55.92002033 | 0.999529746 | 0.999900692 |
| library/737*492, 93KB | 0.0835 | 0.16685455 | 55.90742308 | 0.999590449 | 0.99995085 |
| Gardun/800*532, 123KB | 0.0824 | 0.166457556 | 55.91776847 | 0.999512826 | 0.999937976 |

Table 2: Image Quality Measurements Based on # 4-Significant Bit (4-SB)

| Images/Size/ JPEG | MAE | MSE | PSNR | SSIM | UIQI |
|---|---|---|---|---|---|
| Mug/800*532, 31KB | 0.1664 | 0.666554726 | 49.89244549 | 0.994976996 | 0.999243778 |
| Lane/800*532, 33.7KB | 0.1671 | 0.663404851 | 49.91301718 | 0.992629609 | 0.998838349 |
| Sea/800*532, 57KB | 0.1648 | 0.66324005 | 49.91409618 | 0.998210627 | 0.999603839 |
| library/737*492, 93KB | 0.1653 | 0.664903457 | 49.9032177 | 0.998542376 | 0.999803962 |
| Gardun/800*532, 123KB | 0.1627 | 0.663383085 | 49.91315967 | 0.998118126 | 0.999752758 |

Table 3: Image Quality Measurements Based on Most Significant Bit (MSB)

| Images/Size/ JPEG | MAE | MSE | PSNR | SSIM | UIQI |
|---|---|---|---|---|---|
| Mug/800*532, 31KB | 0.3435 | 2.681579602 | 43.84689667 | 0.981208471 | 0.996961411 |
| Lane/800*532, 33.7KB | 0.338 | 2.682263682 | 43.84578892 | 0.972259217 | 0.995314413 |
| Sea/800*532, 57KB | 0.3356 | 2.668532338 | 43.86807891 | 0.992819986 | 0.99840746 |
| library/737*492, 93KB | 0.3385 | 2.684631171 | 43.84195732 | 0.993361917 | 0.99920829 |
| Gardun/800*532, 123KB | 0.3266 | 2.666044776 | 43.87212922 | 0.992400712 | 0.999005641 |

On the issue of measuring the quality of reconstruction of lossy compression codecs, the PSNR evaluation results shows an interesting findings where the proposed approach achieved notable superiority compared with Self-Generated Verification Code (SGVC) [30] and Multiple Watermarks (MW )[32] as depicted in Table 4 and Figure 3. Another interesting result is on the cumulative squared error between the watermarked and the original image,the Mean Square Error (MSE) result proved also the superiority of the proposed approach compared to Self-Generated Verification Code (SGVC) [30] and Multiple Watermarks (MW )[32] as depicted in Table 4 and Figure 4. Moreover,regarding perception-based model that considers image degradation as perceived change in structural information such as luminance masking and contrast masking, the proposed approach achieved superior results compared to elf-Generated Verification Code (SGVC) [30] and Multiple Watermarks (MW )[32] as depicted in Table 4 and Figure 5.

Table 4: Comparison results based on MAE,MSE,PSNR,SSIM, and UIQI metrics

| Image Quality Metric (IQM) | SGVC [30] | MW[32] | Proposed Approach |
|---|---|---|---|
| AVG-PSNR | 52.16478 | 42.34178 | 55.91465308 |
| AVG-MSE | 0.57906 | 0.97638 | 0.166577359 |
| AVG-SSIM | 0.95148 | 0.92584 | 0.999078513 |
| AVG-MAE | Not Evaluated | Not Evaluated | 0.083816667 |
| AVG-UIQI | Not Evaluated | Not Evaluated | 0.999873788 |

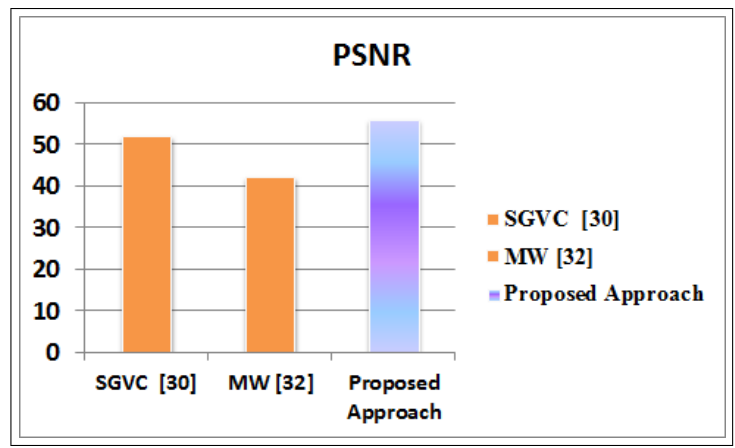

Fig. 3: PSNR Comparison Results

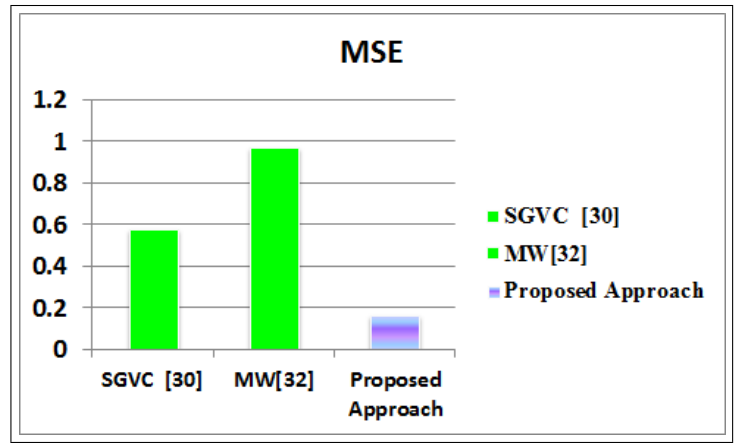

Fig. 4: MSE Comparison Results

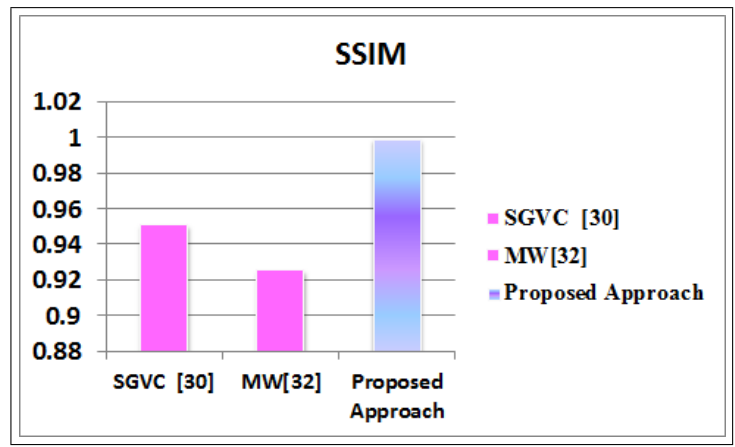

Fig. 5: SSIM Comparison Results

## 4   Conclusion

This study introduced a novel hybrid security approach for protecting colored images against forgery and pixel tampering. The proposed approach is based on Message Digest hashing algorithm (MD5), Advanced Encryption Standard-128 bits (AES), and Stenography. The study has identified the impact of embedding the protection code on image quality measures. The experimental results proved the efficiency of the proposed approach as a novel protection technique for colored images compared with other technique in the literature. This study lays the groundwork for future research into developing more sensitive image forgery detector system and evaluating its performance on different types of colored images and videos.